\documentclass[12pt,twocolumn,letterpaper]{article}

\usepackage[final]{cvpr}              

\usepackage{graphicx}
\usepackage{amsmath}
\usepackage{amssymb}
\usepackage{booktabs}
\usepackage{float}

\usepackage[pagebackref,breaklinks,colorlinks]{hyperref}

\usepackage[capitalize]{cleveref}
\crefname{section}{Sec.}{Secs.}
\Crefname{section}{Section}{Sections}
\Crefname{table}{Table}{Tables}
\crefname{table}{Tab.}{Tabs.}

\def\cvprPaperID{*****} 
\def\confName{CVPR}
\def\confYear{2022}

\begin{document}

\title{COVID-Net US-X: Enhanced Deep Neural Network for Detection of COVID-19 Patient Cases from Convex Ultrasound Imaging Through Extended Linear-Convex Ultrasound Augmentation Learning}

\author{E. Zhixuan Zeng\\
University of Waterloo\\
{\tt\small ezzeng@uwaterloo.ca}
\and
Adrian Florea\\
McGill University\\
St. Mary’s Hospital \\
{\tt\small adrian.florea@mail.mcgill.ca}
\and
Alexander Wong\\
University of Waterloo\\
Waterloo AI Institute\\
DarwinAI Corp.\\
{\tt\small alexander.wong@uwaterloo.ca}
}

\maketitle

\begin{abstract}
As the global population continues to face significant negative impact by the on-going COVID-19 pandemic, there has been an increasing usage of point-of-care ultrasound (POCUS) imaging as a low-cost and effective imaging modality of choice in the COVID-19 clinical workflow.  A major barrier with widespread adoption of POCUS in the COVID-19 clinical workflow is the scarcity of expert clinicians that can interpret POCUS examinations, leading to considerable interest in deep learning-driven clinical decision support systems to tackle this challenge.   A major challenge to building deep neural networks for COVID-19 screening using POCUS is the heterogeneity in the types of probes used to capture ultrasound images (e.g., convex vs. linear probes), which can lead to very different visual appearances.  In this study, we explore the impact of leveraging extended linear-convex ultrasound augmentation learning on producing enhanced deep neural networks for COVID-19 assessment, where we conduct data augmentation on convex probe data alongside linear probe data that have been transformed to better resemble convex probe data.  Experimental results using an efficient deep columnar anti-aliased convolutional neural network designed via a machined-driven design exploration strategy (which we name COVID-Net US-X) show that the proposed extended linear-convex ultrasound augmentation learning significantly increases performance, with a gain of 5.1\% in test accuracy and 13.6\% in AUC.
\end{abstract}

\section{Introduction}
\begin{figure*}[t]
    \centering
    \includegraphics[width=\linewidth]{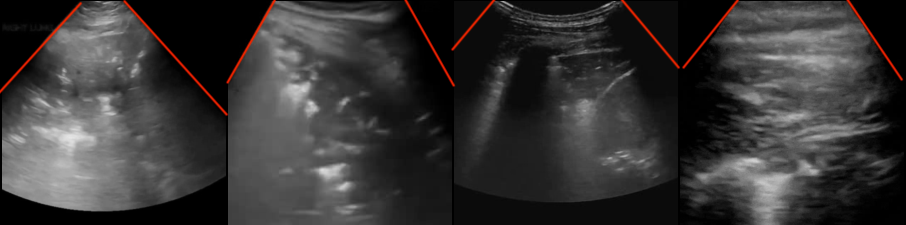}
    \caption{An example of ultrasound images with different viewing windows. The bounds of the viewing windows are marked in red.}
    \label{fig:base_examples}
\end{figure*}

The on-going COVID-19 pandemic, especially with the recent rise of the Omicron variants, has resulted in significant negative impacts on the health and well-being of the global population.  A major aspect of the COVID-19 clinical workflow is imaging-based assessment, and there has been a significant increase in the use of point-of-care ultrasound (POCUS) imaging alongside chest radiography due to its low cost, low maintenance, disinfection ease, portability, and lack of ionizing radiation.  However, a major barrier with widespread adoption of POCUS in the COVID-19 clinical workflow is the scarcity of expert clinicians that can interpret POCUS examinations for COVID-19 assessment \cite{brattain2018machine}.  Therefore, there has been a considerable interest in deep learning-driven clinical decision support systems \cite{maclean2021covid}.

Despite the promise, a major challenge to building deep neural networks for COVID-19 screening is the heterogeneity in the types of probes used to capture ultrasound images.  For example,  POCUS images captured using convex probes have very different visual appearances than images captured using linear probes. Limiting data to a single type increases consistency at the cost of significantly reduced dataset size.

Motivated to tackle this challenge, in this study we explore the impact of leveraging extended linear-convex ultrasound augmentation learning on producing enhanced deep neural networks for COVID-19 assessment.  More specifically,
we leverage projective transformations as part of the data augmentation pipeline to transform linear probe data to better resemble convex data, as well as increase the diversity of viewing windows to improve generalization.  Finally, we will explore its efficacy using  an efficient deep columnar anti-aliased convolutional neural network that provides balance between efficiency and representation capacity.

This paper is organized as follows.  Section 2 describes the methodology behind the proposed extended linear-convex ultrasound augmentation learning strategy for producing enhanced deep neural networks for COVID-19 assessment, as well as the COVID-Net US-X efficient deep columnar anti-aliased convolutional neural network architecture design for exploring the impact of extended linear-convex ultrasound augmentation learning in a clinical resource-constrained scenario.  Section 3 describes the experimental setup used in this study, as well as a discussion of the experimental results studying the impact of the proposed learning strategy.  Finally, conclusions are drawn and future work is discussed in Section 4.

\section{Methods}
\begin{figure}
\centering
    \includegraphics[width=\linewidth]{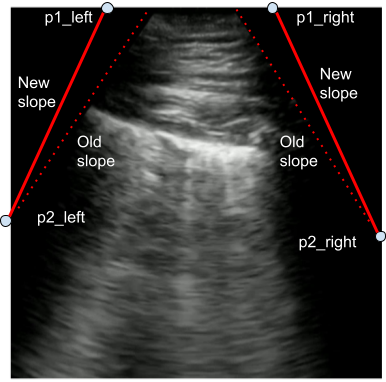}
    \caption{To generate a random transformation: 1. Using the bounds of an ultrasound image defined by the corner points $\{p1_\text{left}, p2_\text{left}, p1_\text{right}, p2_\text{right}\}$, find a new slope based on the distribution $N(\text{old\_slope}, \sigma)$. 2. Define new points using the new slope. 3. Estimate a transformation matrix based on the corner points}
    \label{fig:new_corner_points}
\end{figure}

\begin{figure}
    \begin{subfigure}{0.5\textwidth}
    \includegraphics[width=0.48\linewidth]{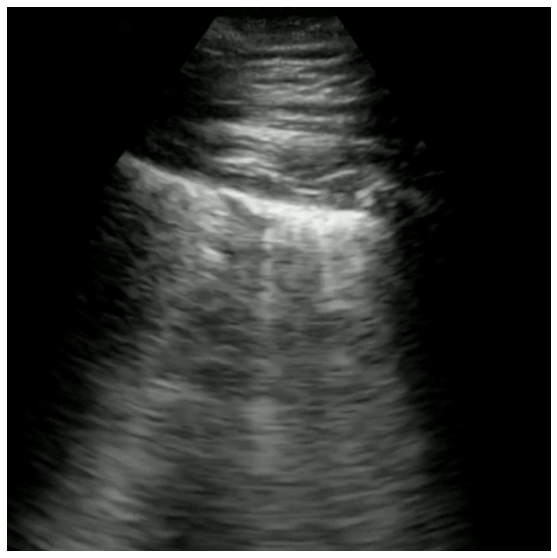}
    \label{fig:convex_base}
        \includegraphics[width=0.48\linewidth]{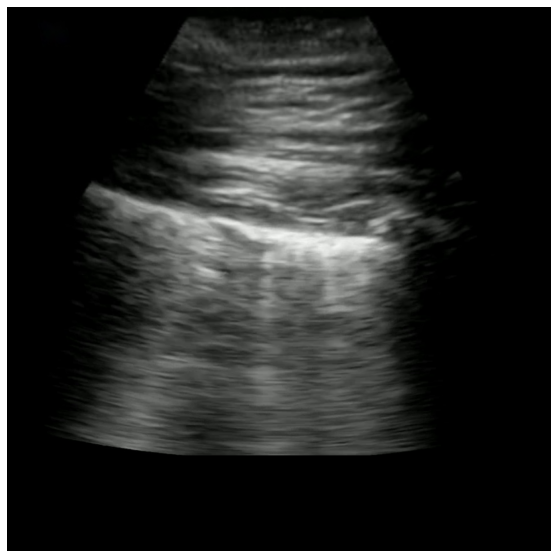}
    \label{fig:convex_proj}     
        \caption{(left) Convex ultrasound image; (right) Augmentation}
    \end{subfigure}
       \\
    \begin{subfigure}{0.5\textwidth}    
    \includegraphics[width=0.48\linewidth]{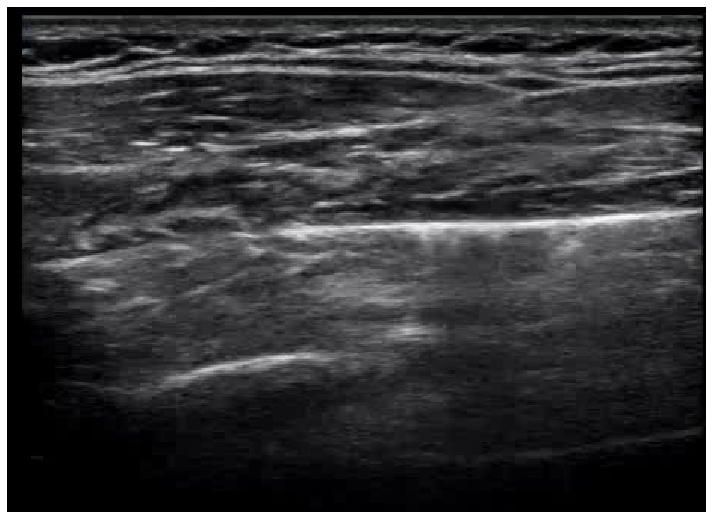}
    \label{fig:linear_base}
    \includegraphics[width=0.48\linewidth]{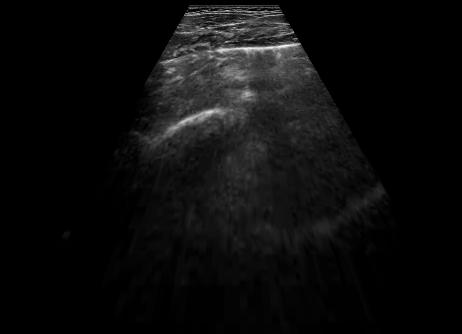}
    \label{fig:linear_proj}
    \caption{(left) Linear ultrasound image; (right) Augmentation}
    \end{subfigure}       
    \caption{Projective data augmentation on convex ultrasound images and linear ultrasound images}
    \label{fig:projective}
\end{figure}

In this study, the concept of extended linear-convex ultrasound augmentation learning is premised on the fact that we wish to leverage as much POCUS data possible to mitigate data scarcity challenges, while at the same time accounting for the heterogeneity in the types of probes used to capture POCUS images that can lead to very significant visual appearance differences.

\begin{figure*}[t]
    \centering
    \includegraphics[width=\linewidth]{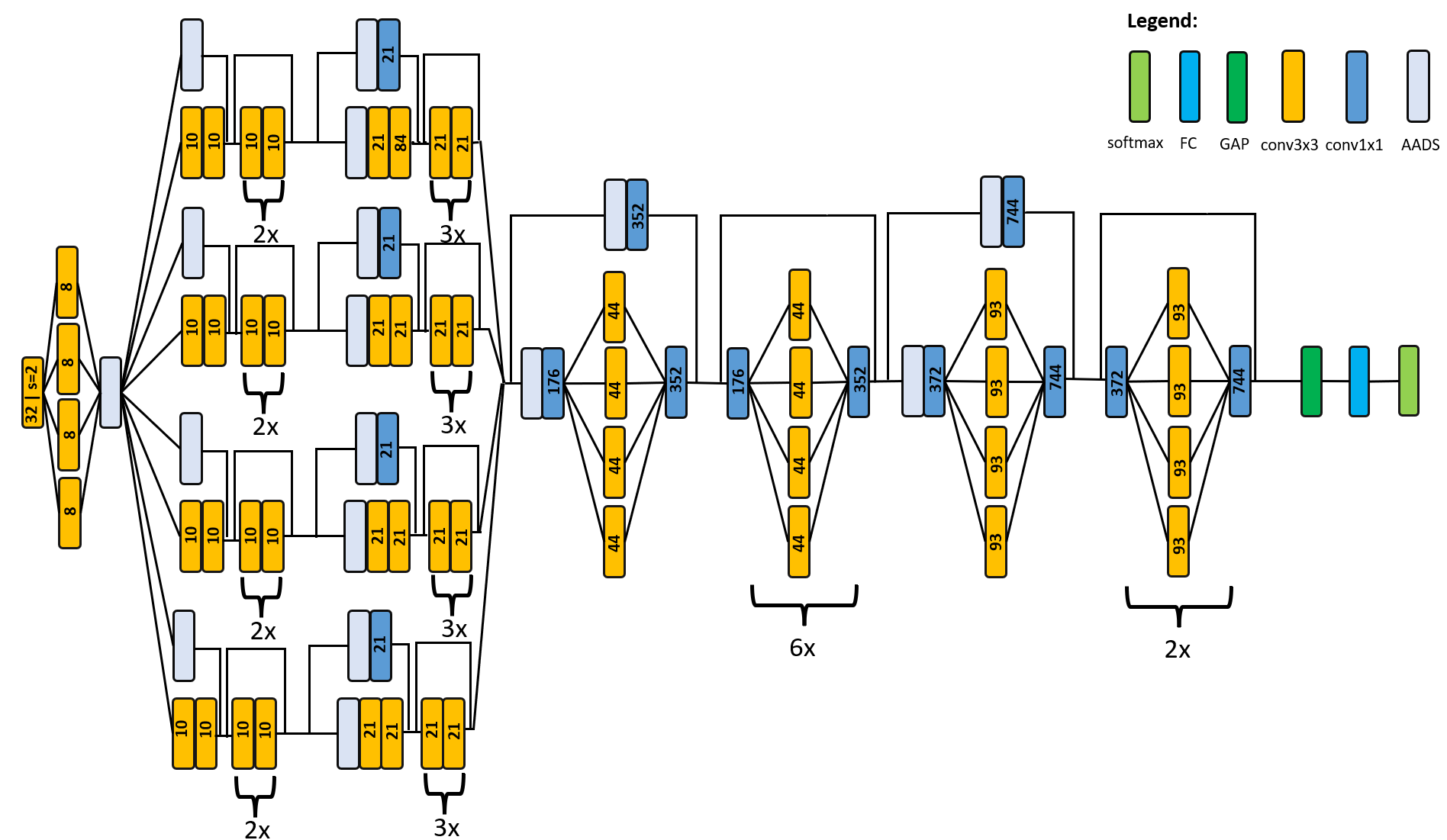}
    \caption{COVID-Net US-X efficient neural network architecture via machine-driven design exploration~\cite{wong2018ferminets}, exhibiting a columnar design with anti-aliasing components.}
    \label{fig:covidnet}
\end{figure*}

For example, POCUS images captured using a convex probe have a greater field-of-view with a cone shaped viewing window. This window differs in its viewing angle from device to device, and can also have a variety of forms as shown in Figure \ref{fig:base_examples}.  In contrast, POCUS images captured using a linear probe has a more restricted field-of-view, but have a linear viewing window with visual content appearing distorted compared to that of images captured using a convex probe.

\subsection{Extended linear-convex ultrasound augmentation learning}
To deal with these various challenges,
we leverage random projective augmentations to not only increase the diversity of viewing windows but also to transform linear probe data in such a way that they better resemble the visual appearance of convex probe data. This will allow for augmented learning with both convex probe data and linear probe data in a way that better promotes generalization since the deep neural network is being exposed to a greater diversity of data but also in a form that is more consistent from visual appearance perspective. \cite{yaron2021point} applies a similar idea through rectifying convex images to appear more similar to linear ones using its polar coordinates, but this process is unable to be used as an on-the-fly data augmentation process

If the bounds of the viewing window is known, then the POCUS image can be transformed to have various different angles by remapping the corners of the viewing window via a random projective transform. This process is shown in Figure \ref{fig:new_corner_points}. Furthermore, we transform the POCUS images captured using linear probes using random projective transforms so that they bare a much stronger visual resemblance to POCUS images captured using a convex probe to enable the deep neural network to better leverage this additional data source.  An example of this POCUS augmentation is shown in Figure 3a and Figure 3b.

\subsection{COVID-Net US-X Architecture Design}
To explore the impact of extended
linear-convex ultrasound augmentation learning in a clinical resource-constrained scenario, we introduce COVID-Net US-X, a convolutional neural network for COVID-19 assessment using POCUS that balances efficiency and representation capacity.  The architecture design is a deep columnar antialiased residual architecture with highly customized macroarchitecture and microarchitecture designs that were automatically determined using a machine-driven design exploration strategy called Generative Synthesis~\cite{wong2018ferminets} for high computational and representational efficiency.  The architecture design has 636M FLOPs and 4.69M parameters.

The advantage of ultrasound imaging systems lies in being a low-cost and low-power solution. As such, using generative synthesis allows us to maintain that power and computational efficiency through automatically striking a balance between detection accuracy and network performance.

The COVID-Net US-X network architecture design features progressively increasing inter-column connectivity as we go deeper into the network.  Such variations in inter-column connectivity enables greater efficiency in the lower degrees of abstraction while achieving greater representational complexity at higher degrees of abstraction.
Futhermore, the COVID-Net US-X network architecture exhibits antialiasing architectural characteristics, which enables greater representational stability to allow the network to be able to achieve greater robustness and greater generalization \cite{zhang2019making}, which is especially important for reliable clinical decision support applications. 
\section{Experiments and Results}
\begin{figure*}
    \centering
    \includegraphics[width=0.6\linewidth]{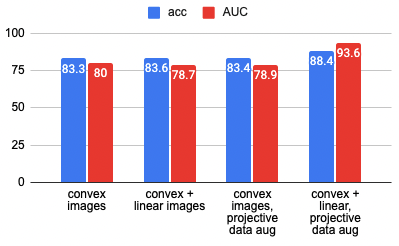}
    \caption{Test accuracy and AUC for classifying COVID-19 on convex and linear ultrasound images with different augmentation strategies (proposed linear-convex augmentation on the far right).}
    \label{fig:results}
\end{figure*}

To evaluate the impact of the proposed extended linear-convex ultrasound augmentation learning strategy, we leveraged a newer version of the COVIDx-US benchmark dataset \cite{ebadi2021covidx} in this study.  To evaluate the impact of the proposed extended linear-convex ultrasound augmentation learning strategy, we leveraged a newer version of the COVIDx-US benchmark dataset \cite{ebadi2021covidx} in this study.  This version of the benchmark dataset consists of 16,649 POCUS images, with 12,260 captured using convex probes and 4389 captured using linear probes, which we leverage fully in this study and is larger and more diverse than what was utilized in~\cite{maclean2021covid}.  In terms of disease breakdown, 6764 of the POCUS images were from COVID-19 positive patients, 9985 were from COVID-19 negative patients. The train, validation, and testing splits are 72\%, 14\%, and 14\%, respectively.

The models are trained with no additional data augmentation techniques. Projective image transformations are set with a slope sampled from a normal distribution centered around either the original slope for convex images or 2.5 for linear images, and a standard deviation of 0.15.

The test accuracy and area-under-the-curve (AUC) for the tested deep neural network under different forms of data augmentation is shown in Fig. [\ref{fig:results}].  It can be observed that the proposed extended linear-convex ultrasound augmentation learning strategy significantly increases performance, with a gain of 5.1\% in test accuracy and 13.6\% in AUC.  Therefore, these results clearly illustrate that the proposed augmented learning strategy can have significant benefits in leveraging additional data from different probe types to improve the performance and generalization of deep neural networks for COVID-19 assessment using POCUS images.

\section{Conclusions}

In this study, we introduced an extended linear-convex ultrasound augmentation learning strategy for creating enhanced deep neural networks for COVID-19 assessment using POCUS images.  More specifically, we performed data augmentation on convex probe data alongside linear probe data randomly transformed in a way that better resembles convex probe data.  We conducted experiments using COVID-Net US-X, an efficient deep columnar anti-aliased convolutional neural network designed via a machined-driven design exploration strategy that we introduced here to evaluate the impact of the proposed augmentation learning strategy. Experimental results demonstrated the efficacy of the proposed extended linear-convex ultrasound augmentation learning in significantly increasing performance, with a gain of 5.1\% in test accuracy and 13.6\% in AUC.

Future work involves investigating the efficacy of the proposed extended linear-convex ultrasound augmentation learning strategy on a larger, more diverse set of POCUS images.  Furthermore, we plan on exploring additional ultraound-specific augmentation types as well as different augmentation ranges to determine the optimal augmentation learning policy for building enhanced deep neural networks for COVID-19 assessment via POCUS images.

{\small
\bibliographystyle{ieee_fullname}
\bibliography{egbib}
}

\end{document}